\documentclass[12pt]{article}
\usepackage{latexsym, epsfig, graphics}
\newcommand{\be} {\begin{equation}}
\newcommand{\ee} {\end{equation}}
\newcommand{\bq} {\begin{eqnarray}}
\newcommand{\eq} {\end{eqnarray}}
\begin{document}
\begin{titlepage}
\today          \hfill 
\begin{center}
\hfill    LBNL-54837\\
          \hfill    UCB-PTH-04/10\\

\vskip .5in

{\large \bf Meanfield Approximation For Field Theories
On The Worldsheet Revisited}
\footnote{This work was supported in part
 by the Director, Office of Science,
 Office of High Energy and Nuclear Physics, 
 of the U.S. Department of Energy under Contract 
DE-AC03-76SF00098, and in part by the National Science Foundation Grant
PHY-0098840}
\vskip .50in


\vskip .5in
Korkut Bardakci\footnote{e-mail: kbardakci@lbl.gov}

{\em Department of Physics\\
University of California at Berkeley\\
   and\\
 Theoretical Physics Group\\
    Lawrence Berkeley National Laboratory\\
      University of California\\
    Berkeley, California 94720}
\end{center}

\vskip .5in

\begin{abstract}
This work is the continuation of the earlier efforts to apply
the mean field approximation to the continuum world sheet formulation
of planar $\phi^{3}$ theory. The previous attempts were either
simple but without solid foundation or better founded but
excessively complicated. In this article, we present an approach
both simple, and also systematic and well founded. We are able
to carry through the leading order mean field calculation
analytically, and with a suitable tuning of the coupling
constant, we find string formation.

\end{abstract}
\end{titlepage}

\newpage
\renewcommand{\thepage}{\arabic{page}}
\setcounter{page}{1}
\noindent{\bf 1. Introduction}

\vskip 9pt

Over the last couple of years, the present author and Charles Thorn
have been pursuing a program of summation of planar graphs in
field theory [1,2,3,4]. Because of its simplicity, the field
theory that has been most intensively investigated so far is
the $\phi^{3}$ theory, although Thorn and collaborators have made
considerable progress in extending the program to more
physical theories [5,6,7]. The basic idea, which goes back to
't Hooft [8], is to represent  Feynman graphs on the world
sheet by a suitable choice of the light cone coordinates. 't Hooft's
representation, which was non-local, was later reformulated as a
local field theory on the world sheet by introducing additional
non-dynamical fields [1]. This reformulation has no new physical
content; it merely reproduces perturbation theory. However, it
provides a setup well suited for the study of string
formation in field theory. This is an old problem that has
attracted recent renewed interest [9,10,11,12] following the discovery of
AdS/CFT correspondence [13,14].

Our approach to the problem of string formation starts with the
world sheet description of the $\phi^{3}$ field theory mentioned above,
and we look for the phenomenon of ``condensation'' of Feynman graphs.
This phenomenon will be defined more precisely later in the paper,
but roughly it means that the lines that form Feynman graphs on
the world sheet become dense, and graphs of
asymptotically infinite order dominate the perturbative expansion.
Furthermore, the original non-dynamical
world sheet fields become dynamical and string
formation takes place.
Whether the scenario described above really happens is of course
a question of dynamics. So far, the only tool used to investigate
this problem in the present context
 has been the mean field or the self consistent field
approximation [2,3,4]. The accuracy of the mean field
approximation is questionable; however,
 one can hope that at least a qualitative understanding
of the relevant dynamical issues would emerge.

The main virtue of the mean field method is its simplicity.
There is, however, no unique way to do the mean field
calculation, and it all depends on the choice of the order
parameter. In reference [2], the simplest choice was made
for the order parameter by taking it
to be the expectation value of a local field. This makes
the subsequent calculation quite tractable. However, this
early attempt, at least in its continuum version, relied
on a number of questionable steps and approximations that are
hard to justify.
 In order to overcome these difficulties, in references
[3] and [4] the order parameter was taken to be the expectation
value of two fields at different points (two point function). This also
has the advantage of providing a better probe of the problem; but the
disadvantage is that the calculation becomes too complicated
to carry out analytically.

In this article, we revisit the earlier calculation [2] based
on a simple order parameter to see whether the difficulties
associated with it can be overcome without sacrificing its
basic simplicity. We will mainly focus on the treatment based
on the continuum world sheet given in this reference.
 Reexamining this treatment,
we identify the following problems:\\
a) The boundary conditions on the world sheet were imposed
only approximately through the so called $\beta$ trick. The
exact boundary conditions could only be recovered in the
 problematic $\beta\rightarrow\infty$ limit.\\
b) If we try to impose the boundary conditions exactly by
means of Lagrange multipliers, as was done in [1] and as we
shall do here, we avoid the problem discussed in a), but
instead we encounter another difficulty. The action has then an
unfixed gauge invariance, which can lead to ill defined 
results.\\
c) There are two kinds of fields in the problem: The matter and
the ghost fields. The contribution of each sector is quadratically
divergent, but there is a subtle partial cancellation between them.
Unless great care is exercised, the result can depend on
the regulation scheme used.\\
d) The use of light cone coordinates obscures the covariance
of the theory. An approximation scheme, such as the mean field
approximation, could easily violate Lorentz invariance.

In this article, we propose the following improvements to
overcome the problems listed above:\\
a) The boundary conditions will be imposed exactly
by means of Lagrange multipliers.\\
b) To avoid the resulting gauge invariance problem,
we introduce a gauge fixing term. This is probably
the most important new idea
 introduced in the present work. As is well known,
working with a gauge unfixed action can lead to
lots of confusion.\\
c) To keep track of the cancellation between the mattter
and ghost fields, we impose supersymmetry on the
world sheet. This idea was already in the air in [1],
what we have done here is to formulate it explicitly.
 Also, in treating divergent quantities, 
 we first combine the contributions
of the matter and ghost sectors, and then we regulate the
sum in a way that does not spoil the cancellation
between them. The answer obtained in this fashion is
unambiguous.\\
d) There is a particular subgroup of the Lorentz group
under which the light cone variables transform linearly.
In particular, under boosts along the $1$ direction,
the variables $x^{\pm}$ and $p^{\pm}$ scale (see section
2 for the definition of these variables). All of this
is familiar from the light cone treatment of the bosonic
string, where, among other things, the importance of 
invariance under this special boost was recognized [15].
In the present context, this was discussed in reference
[4]. In this paper,  following the treatment given
in [4], we will try to preserve invariance under this
special subgroup at each step. We will not, however,
try to investigate invariance under the full Lorentz group.
In string theory, full Lorentz invariance in the light cone
framework is realized only for a fixed critical value
of $D$, where $D$ is the dimension of the transverse space
 [15]. If the same is true here, this problem
is clearly beyond the scope of a leading large $D$
approximation.

When these improvements are incorporated in the
mean field approximation of reference [2], the result
is a systematic approximation scheme without any of
the ambiguities encountered in [2]. It is also simple
enough so that we are able to carry out the leading
order calculation analytically. With a suitable tuning
 of the coupling  constant, we find string formation,
and in addition, we discover that a new string mode
has been dynamically generated.

In organizing this paper, the goal was to present a
complete and self-contained treatment which should
be intelligible even to a reader unfamiliar with
the previous work on the subject. As a result, there
a good deal of overlap with reference [2], and some
overlap with references [3] and [4]. When we preview
the rest of the paper below, we will try to make
clear what is new and what is a review or eloboration
of the material covered in the earlier work cited
above.

In section 2, we briefly review both the rules
for Feynman graphs in light cone variables [8] and 
the local field theory on the world sheet from which
these graphs follow [1]. We also discuss the transformation
properties of the  fields under the special
boost mentioned earlier, which manifests itself as a
scale transformation on the world sheet. This is
an abridged version of a more complete discussion 
given in [4].

In section 3, world sheet supersymmetry is introduced
and a supersymmetric free action $S_{0}$ is constructed.
This is a new idea in the present context.
$S_{0}$ differs from the corresponding action introduced
in [1] and used in all the subsequent work in the
structure of the terms involving ghosts, and also in
the presence of a relatively insignificant bosonic
field ${\bf r}$ required by SUSY.
 The cancellation
of quantum corrections between matter and ghost fields,
which was the reason for the introduction of ghosts
in the first place, is now guaranteed by supersymmetry.

The boundary conditions accompanying $S_{0}$ are enforced
by a term in the action, $S_{1}$, given in section 4. Both
the boundary conditions and $S_{1}$ are substantially the 
same as those in [1]. To express the new term in the
action fully in the language of field theory, one needs
world sheet fermions which were introduced earlier [1,4].
 We have found it slightly more convenient to work with
 a somewhat different set of fermions,
 although it is
easy to show by means of a Bogoliubov transformation that
the two are completely equivalent.

In section 5, we show that the action constructed so far
is invariant under a  simple gauge transformation.
It is therefore important to fix this gauge, and we show how
to do it. We also note that it is possible to introduce
some arbitrary constants in the boundary and gauge fixing
terms in the action. Although the exact theory is not
sensitive to these constants, we keep them around
to see how the affect an approximate calculation. The
material covered in this section is completely new.

The mean field approximation which is at the basis of the 
present work is discussed in section 6 from
the point of view of the large $D$ limit. All of this is
standard material, well known from the solution of the
vector models in in the large $N$ (in this case, large
$D$) limit [16]. The only thing new here compared to
[2] and [3] is the manner in which
 the singular determinants resulting from integration
over the matter and ghost fields are regulated: The
two determinants are combined into an expression less
singular than each one seperately, and regulating the
combined expression, we get an  unambiguous and scale 
invariant answer.

In section 7, an effective action is constructed and
evaluated by the saddle point method in the large
$D$ limit. This effective action is pretty close
to but still different in detail
 from the one derived in [2].
 A standard bosonic string action
with positive slope emerges from
this calculation. The important question is then
whether this slope is finite. We find that, by
suitably tuning the coupling constant of the 
$\phi^{3}$ theory, the slope can be made finite.
This tuning can be regarded as renormalization:
A cutoff dependent coupling constant is traded
for the finite slope parameter.

It seems somewhat surprising that starting with
an unstable field theory, no sign of instability
has so far appeared in the string picture.
 One possibility is that we have not
gone far enough. The calculation of the intercept,
which we have not undertaken here, may show that,
as in the bosonic string, some lowest lying
states are tachyonic. Or, it may be that the
instability is not visible in the leading order
of the large $D$ limit. These possibilities are
discussed in at the end of section 7.

So far, all the calculations were carried out
in the leading order of the large $D$ limit.
In section 8, we compute a non-leading correction
in this limit by expanding the composite
field $\rho$ (see section 4 for its definition)
around its mean value $\rho_{0}$. In addition,
$\rho$ is assumed to be slowly varying, and an
expansion up to second order in powers of
derivatives of this field is carried out. This
is essentially a repetition of the computation
done in the Appendix B of [2] from the
standpoint of the present approach. We find that,
from the world sheet perspective, $\rho$ becomes
a dynamical, propagating field, and from the string
perspective, the string acquires an additional mode,
with the same slope as all the other modes. Finally,
in the Appendix, we show that, the mean field
computation presented here is completely equivalent
to the standard large $N$ ($D$) treatment of vector
models.

\vskip 9pt
 
\noindent{\bf 2. A Brief Review}
\vskip 9pt

The Feynman graphs of massles $\phi^{3}$ have a particularly simple
form in the mixed light cone representation of 't Hooft [8]. In this
representation, the evolution parameter is $x^{+}$ also denoted
by $\tau$, and the conjugate Hamiltonian is $p^{-}$, and the
Minkowski evolution operator is given by
\be
T=\exp(-i x^{+} p^{-}).
\ee
In this notation, a Minkowski vector $v^{\mu}$ has the light cone
components $(v^{+},v^{-},{\bf v})$, where $v^{\pm}=(v^{0}\pm v^{1})
/\sqrt{2}$, and the boldface letters label the transverse directions.
A propagator that carries momentum $p$ is pictured as a horizontal 
strip of width $p^{+}$
 and length $\tau=x^{+}$ on the world sheet,
bounded by two solid lines (Fig.1).
\begin{figure}[t]
\centerline{\epsfig{file=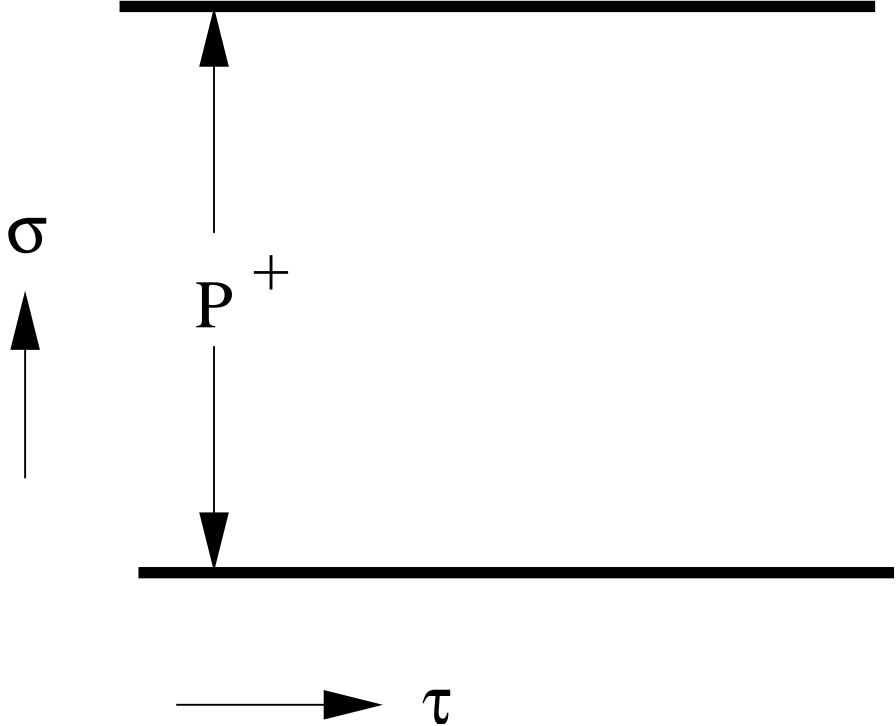,width=12cm}}
\caption{The Propagator}
\end{figure}
 The lines forming the boundary
carry transverse momenta ${\bf q}_{1}$ and ${\bf q}_{2}$, where
$$
{\bf p}={\bf q}_{1}-{\bf q}_{2},
$$
and the corresponding propagator is given by
\be
\Delta(p)=\frac{\theta(\tau)}{2 p^{+}}\exp\left(-i\,\frac{\tau}
{2 p^{+}}\, {\bf p}^{2}\right).
\ee

More complicated graphs consist of several horizontal solid line segments (Fig.2).
\begin{figure}[t]
\centerline{\epsfig{file=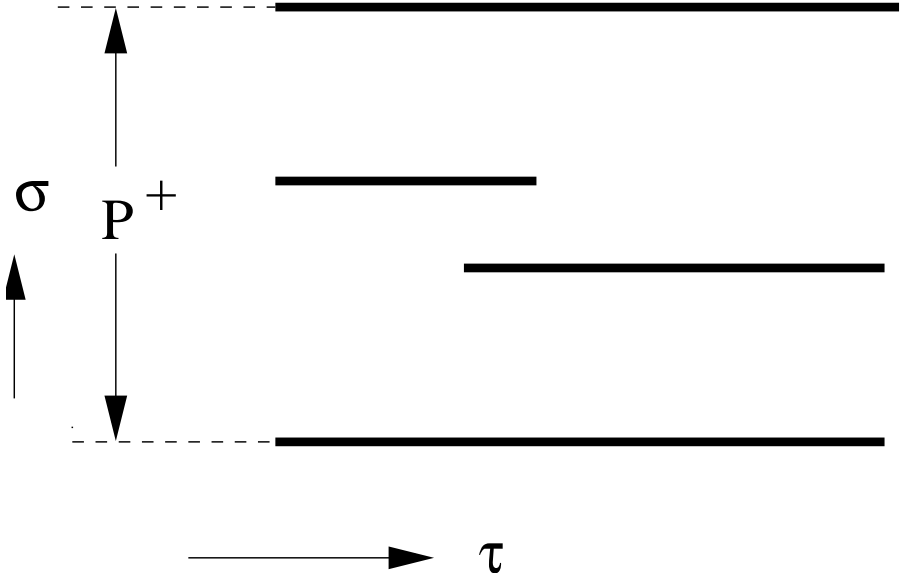,width=12cm}}
\caption{A Typical Graph}
\end{figure}
The beginning and the end of each segment is where the $\phi^{3}$
interaction takes place, and a factor of $g$ is associated with each
such point, where $g$ is the coupling constant. Finally, one has
to integrate over the position of the interaction vertices, as well
as the momenta carried by the solid lines. We note that momentum conservation
is automatic in this formulation. A typical light cone graph is
pictured in Fig.2.

It was shown in [1] that the light cone Feynman rules described 
above can be reproduced by a local field theory on the
world sheet. The world sheet is parametrized by coordinates $\sigma$
along the $p^{+}$ direction
and $\tau$ along the $x^{+}$ direction, and the transverse
momentum ${\bf q}$ is promoted to a bosonic field ${\bf q}(\sigma,
\tau)$ defined everywhere on the world sheet. In addition, two
fermionic fields (ghosts) $b(\sigma,\tau)$ and $c(\sigma,\tau)$
are needed. In contrast to ${\bf q}$, which has D components, $b$
and $c$ each have D/2 components, where D is the dimension of the
transverse space, assumed to be even.
 The action on the world
sheet, with the Minkowski signature (+,-),   is given by
\be
S_{0}=\int_{0}^{p^{+}} d\sigma \int_{\tau_{i}}^{\tau_{f}} d\tau
\left(b'\cdot c'- \frac{1}{2} {\bf q}'^{2}\right),
\ee
where the prime denotes derivative with respect to $\sigma$. This
action is supplemented by the Dirichlet conditions
\be
\dot{{\bf q}}=0,\; b=c=0,
\ee
on the solid lines (boundaries),
where the dot denotes derivative with respect to $\tau$. Imposing
Dirichlet conditions on ${\bf q}$ is equivalent to fixing them to be 
$\tau$ independent  on the solid lines
 and then integrating over them. If we fix ${\bf q}$ to be
${\bf q}_{1}$ and ${\bf q}_{2}$ on two adjacent solid lines and
solve the equations of motion
$$
{\bf q}''=0
$$
subject to these boundary conditions, we find that the the resulting
classical action reproduces the exponential factor in
eq.(2). However, there is also an unwanted
 quantum contribution, given by
$$
- \frac{1}{2} D \det(\partial_{\sigma}^{2})
$$
which is exactly cancelled by the corresponding determinant resulting
from integrating over $b$ and $c$.

The action formulation described above, which was extensively used
in the previous work [2,3,4]. Although it is basically correct, it
 has some unsatisfactory features. For one
thing, although we have nothing new to say about this problem in
this paper,
 the factor of $1/(2 p^{+})$ in front of the exponential in
eq.(2) is missing. Also, the splitting of the ghost fields into two
components $b$ and $c$ is somewhat unnatural; it leads to the 
artificial condition that D is even, and it does not look
 rotationally invariant, although there is
of course no real violation of rotation symmetry in the
transverse space. In the next section,
we will show that the introduction of supersymmetry on the world
sheet leads to a more symmetric ghost sector and the condition that
D be even is no longer needed. We also think that it  results
in a more natural and elegant approach.

Finally, we would like to discuss briefly the question of Lorentz
invariance. This is a non-trivial problem, since the use of the
light cone coordinates obscures the Lorentz transformation properties
of the dynamical variables. There is, however, a special subgroup
of the Lorentz group under which the light cone coordinates have
simple linear transformation properties \footnote{In this article,
we will not investigate invariance under the remaining Lorentz
generators. If the string analogy is to be trusted, this is where
the critical dimension becomes important [15], and a large $D$
approximation is clearly inadequate for treating this problem}.
 If $L_{i,j}$ are the
angular momenta and $K_{i}$ are the boosts, the generators of this
subgroup are
\be
L_{i,j},\;\; M_{+,-}=K_{1}, \;\; M_{+,i}=K_{i}+L_{1,i},
\ee
where indices $i$ and $j$ run from 2 to $D+2$. It turns out
that invariance under all the generators except for one is
rather trivially satisfied by the propagator (2) or the field
theories (3). The non-trivial transformation, generated by
$K_{1}$, corresponds to scaling of
$x^{+}$ and $p^{+}$ by a constant u:
\be
x^{+}\rightarrow x^{+}/u,\;\; p^{+}\rightarrow p^{+}/u,
\ee
and the tranverse momenta ${\bf q}$ are unchanged. The reason
this transformation is critical is that although it is easy
to construct  classically scale invariant theories, this symmetry
is in general broken by quantum corrections. This is 
familiar from the study of conformal invariance in field
theory.

Now, let us examine the scale invariance of the action (3)[4]. Since
the coordinates $\sigma$ and $\tau$ must transform like $p^{+}$
and $x^{+}$, the transformation laws of the fields are
\be
{\bf q}(\sigma,\tau)\rightarrow {\bf q}(u\sigma,u\tau),\;\;
b,c(\sigma,\tau)\rightarrow b,c(u\sigma,u\tau),
\ee
and the classical action is invariant under this transformation.
On the other hand,
the quantum corrections are singular and need a cutoff for
their definition. This cutoff would break scale invariance, were
it not for the cancellation between the ghost and matter fields.
So what we have here is a potential violation of scale invariance
by an anomaly, which is eliminated by the cancellation between
the matter and ghost sectors. 
This is nothing but the cancellation of the determinants discussed 
earlier. We would like to stress that the quantum contribution in
question is a world sheet effect; it potentially present even in the
case of a free propagator and it has nothing to do with the target
space ultraviolet divergences.

In addition to the scaling of the bulk,
 we have to consider the scaling behaviour of the
 boundary conditions given by
eq.(4). These are scale invariant as they stand, but the integration
over the position of the boundaries is not invariant, since
the $\sigma$ coordinate scales. The factor of $1/(2 p^{+})$
provides the measure needed to make this integration scale 
invariant. Although we will not present here a 
general recipe for the inclusion of this factor,
 we will be careful to preserve the scale
invariance of various integrals that occur in the course of
the mean field calculation. It can be shown that, in any case, 
this factor does not contribute to the leading order of the
mean field calculation [2,3].

\vskip 9pt

\noindent{\bf 3. SUSY On The World Sheet}

\vskip 9pt

We generalize the momentum ${\bf q}$ to form a SUSY multiplet:
\be
{\bf Q}={\bf q}+\theta_{1}{\bf b}+\theta_{2} {\bf c} +
\theta_{1} \theta_{2} {\bf r},
\ee
where $\theta_{1,2}$ are the usual anticommuting coordinates
and all the fields represented by boldface letters are vectors
in the D dimensional transverse space. The two SUSY generators
are($i=1,2$)
\be
Q_{i}=\frac{\partial}{\partial \theta_{i}}+ \theta_{i}
\partial_{\sigma},
\ee
and they satisfy
\be
Q_{1}^{2}=Q_{2}^{2}=\partial_{\sigma},\;\;
[Q_{1},Q_{2}]_{+}=0.
\ee
The infinitesimal transformations are given by
\be
{\bf Q}\rightarrow {\bf Q}+[\sum_{i}\epsilon_{i}
Q_{i},{\bf Q}],
\ee
where $\epsilon_{1,2}$ are anticommuting infinitesimal
parameters. We also define two covariant derivatives
\be
D_{i}=\frac{\partial}{\partial \theta_{i}}-\theta_{i}
\partial_{\sigma},
\ee
which anticommute with the SUSY generators and satisfy
the same equations as (10), apart from the change of the
sign of $\partial_{\sigma}$.

Having introduced the framework of supersymmetry, we
write down the supersymmetric analogue of the action
(3):
\bq
S_{0}&\rightarrow&\int_{0}^{p^{+}} d\sigma \int d\tau \int
d^{2}\theta\left(-\frac{1}{2} D_{1}{\bf Q}\cdot
D_{2}{\bf Q}\right)\nonumber\\
&=&\int_{0}^{p^{+}} d\sigma \int d\tau\left(
-\frac{1}{2}({\bf q}'^{2}+{\bf b}\cdot {\bf b}'
+{\bf c}\cdot {\bf c}'+{\bf r}^{2})\right),
\eq
supplemented by the boundary conditions
\be
\dot{{\bf q}}=0,\;\;{\bf b}=0,\;\;{\bf c}=0,
\;\;{\bf r}=0,
\ee
on the solid lines. We now have twice as many ghost
fields as before, but they appear with only first
order  derivative in $\sigma$, and as a result,
 there is again
complete cancellation between the determinants(
quantum corrections)
coming from the integration of the matter and ghost
fields. It is clear that this cancellation is a
consequence of supersymmetry. In fact, the main
motivation for introducing SUSY was to obtain this
 cancellation as a systematic 
conseqence of a symmetry, and not as some accident.
In the rest of this article, we will exclusively use
this supersymmetric form of the action.

Although there is supersymmetry in the bulk of the
worldsheet, it is broken by the boundary conditions,
since the condition on ${\bf q}$ differs from those
on ${\bf b}$ and ${\bf c}$. This breakdown of SUSY
is essential, since supersymmetric boundary conditions
would lead to a complete cancellation between the matter
and the ghost sectors, resulting in a trivial propagator.
This has no effect on the
cancellation of the quantum contributions, since the
cancellation  occurs in the bulk
and it is insensitive to the boundary conditions.

Since we now have a new expression for $S_{0}$, we have
to reinvestigate the invariance under the scaling
transformations (6,7). The action given by eq.(13) is invariant 
if the fields transform as
\bq
{\bf q}(\sigma,\tau)&\rightarrow&{\bf q}(u\sigma,u\tau),
\;\;\;{\bf b}(\sigma,\tau)\rightarrow \sqrt{u}\, {\bf b}
(u\sigma,u\tau),\nonumber\\
{\bf c}(\sigma,\tau)&\rightarrow& \sqrt{u}\, {\bf c}(u\sigma,
u\tau),\;\;\; {\bf r}(\sigma,\tau)\rightarrow u\, {\bf r}
(u\sigma,u\tau).
\eq
Again, a potential quantum anomaly that could violate
scale invariance is cancelled as a consequence of
supersymmetry.

\vskip 9pt

\noindent{\bf 4. The World Sheet Action}

\vskip 9pt

In this section, we review the construction of an action
that incorporates the boundary conditions (14), which will
be implemented  by introducing a bosonic Lagrange multipliers
${\bf y}(\sigma,\tau)$ and ${\bf z}(\sigma,\tau)$,
 and the fermionic Lagrange multipliers
$\bar{{\bf b}}(\sigma,\tau)$ and $\bar{{\bf c}}(\sigma,\tau)$
[1]. The corresponding term in the action is
\be
S_{1}=\int_{0}^{p^{+}} d\sigma\int d\tau\,\rho\left({\bf y}\cdot
\dot{{\bf q}}+ \bar{{\bf b}}\cdot{\bf b}+\bar{{\bf c}}\cdot
{\bf c}+{\bf z}\cdot{\bf r}\right),
\ee
where $\rho(\sigma,\tau)$ is equal to $1$ on the solid lines
and it is equal to $0$ elsewhere. This definition is 
singular on the continuum world sheet; we will give a more precise
definition on the discretized world sheet below. 
 The factor $\rho$ ensures that the
boundary conditions are imposed only on the solid lines; elsewhere,
$\rho$ vanishes and there are no constraints on the fields.

In the earlier work [1,2,4], it was shown how to construct $\rho$ in
terms of a fermionic field. We will present here a slightly
different version of this construction.
In order to avoid singular expressions, it is best to start by
 discretizing the $\sigma$ coordinate into segments of length
$\Delta\sigma=a$, with $p^{+}=N a$. 
 The parameter $a$
plays the role of a cutoff, which will in any case be needed
later on. The specific form
of this cutoff is not important; for example, a cutoff in the Fourier
modes conjugate to $\sigma$ would serve just as well.
Fig.3 shows $N$ equally spaced lines parallel to the $\tau$ direction.
\begin{figure}[t]
\centerline{\epsfig{file=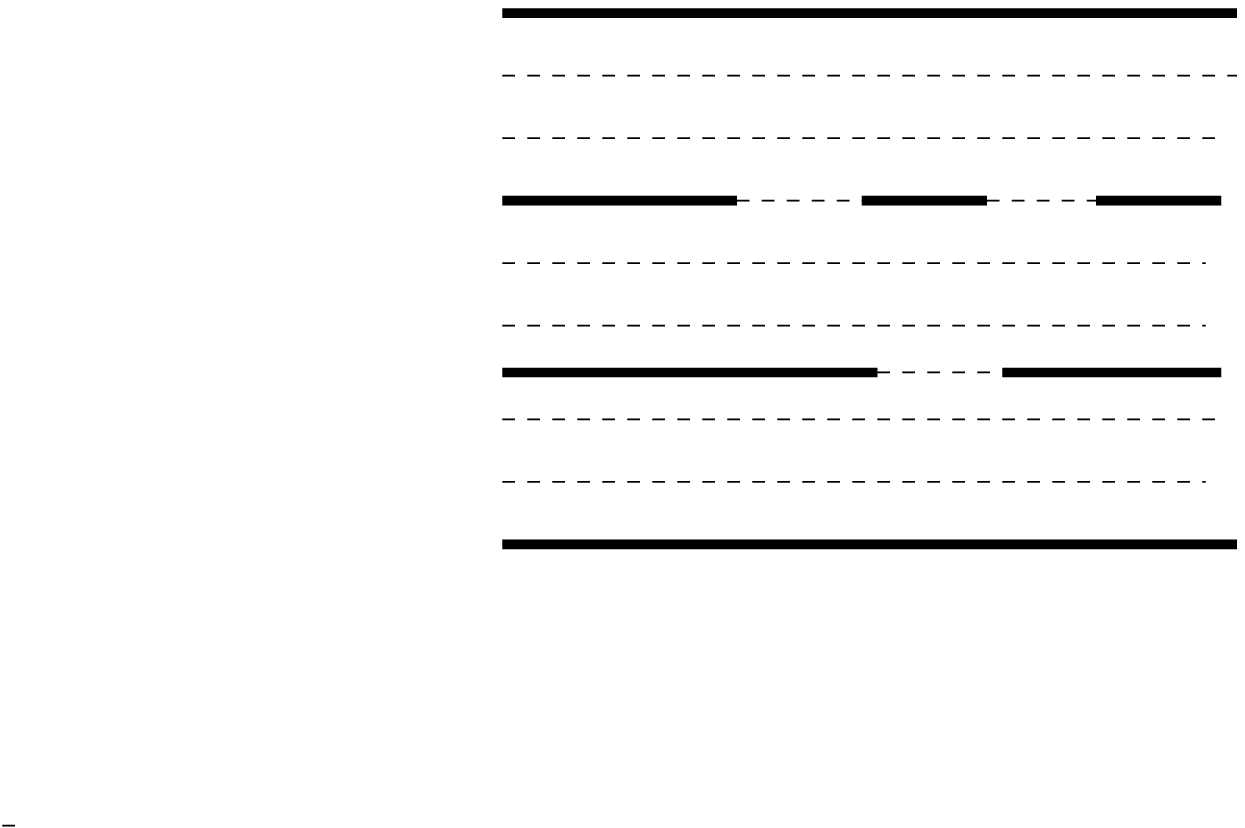,width=12cm}}
\caption{Solid And Dotted Lines}
\end{figure}
For the convenience of exposition, the absence of a solid line in
a given position is pictured by the presence of a dotted line
in the same position. In other words, the solid lines mark the
boundaries and the dotted lines fill the bulk of the world sheet.
 On each line, we introduce two fermionic
variables $\psi_{i}(\sigma_{n},\tau)$ and their conjugate
$\bar{\psi}_{i}(\sigma_{n},\tau)$, where $i=1,2$,
$\sigma_{n}=n a$, and $n$ is
an integer in the range $0\leq n\leq N$.  The free fermionic
action is given by
\be
S_{2}=\sum_{n}\int d\tau\, i\bar{\psi}(\sigma_{n},\tau) \dot{\psi}
(\sigma_{n},\tau),
\ee
and the fermions satisfy the usual anticommutation relations:
$$
[\bar{\psi}_{i}(\sigma_{m},\tau),\psi_{j}(\sigma_{n},\tau)]_{+}
=\delta_{i,j}\, \delta_{m,n}.
$$
The function of the fermions is to keep track of the solid and 
dotted lines. The vacuum state, defined by
$$
\psi_{i}|0\rangle =0,
$$
corresponds to the trivial situation where all the lines in the graph
are dotted. The $\tau$ independent
state corresponding to a single solid line at $\sigma=
\sigma_{n}$ is
\be
|\sigma_{n}\rangle=\bar{\psi}_{1}(\sigma_{n},\tau)\bar{\psi}_{2}
(\sigma_{n},\tau)|0\rangle.
\ee
This is an eternal solid line, extending indefinitely for both positive
and negative $\tau$, since
the operators $\bar{\psi}$, and hence the above state, are independent
of $\tau$ by the equation of motion following from (17). Several solid
lines are represented by the state
$$
\prod_{n}\bar{\psi}_{1}(\sigma_{n})\bar{\psi}_{2}(\sigma_{n})|0\rangle.
$$
We note that, a state with a double solid line, which has no graphical
interpretation, vanishes by fermi statistics.
Having set up our fermionic system, we can express $\rho$ in eq.(16) in terms
of fermions:
\be
\rho(\sigma_{n},\tau)=\frac{1}{2}\sum_{i=1,2} \bar{\psi}_{i}(\sigma_{n},
\tau)\psi_{i}(\sigma_{n},\tau),
\ee
and it is easy to check that $\rho$ is $1$ if there is a solid line
located at $\sigma=\sigma_{n}$, and it is zero if the line is dotted.
The ground state expectation value of
this composite field, $\rho_{0}$, will play an important role in what follows.
In any finite order of perturbation theory, where the density of solid
lines is essentially zero, $\rho_{0}$ is zero. On the other hand, a
non-zero $\rho_{0}$ means that a portion of the world sheet with finite
area is occupied by the solid lines, which we interpret  as condensation
of solid lines. It is also clear that this can only happen if large (infinite)
order graphs dominate the perturbation series. It will be shown later that,
at least in the mean field approximation, the condensation of solid
lines leads to string formation. One can think of $\rho_{0}$ as an
order parameter that distinguishes between the stringy phase and the
perturbative phase of the same theory. 

So far, all the lines, whether solid or dotted, are eternal. We need an 
interaction term in the action which will convert dotted lines into solid
lines and vice versa. Remembering that the transition between solid and
dotted lines is accompanied by a factor of the coupling constant $g$,
we set
\be
S_{3}= g \sum_{n}\int d\tau\left(\bar{\psi}_{1}(\sigma_{n},\tau)
\bar{\psi}_{2}(\sigma_{n},\tau)+\psi_{2}(\sigma_{n},\tau)
\psi_{1}(\sigma_{n},\tau)\right),
\ee
and it is easy to show that this term in the action does the
required job.

As we mentioned earlier, the fermions introduced in this paper are
somewhat different from those used in the earlier work [1,2,4]. However,
it is not difficult to show that the two
 are connected by the Bogoliubov transformation
$$
\bar{\psi}_{1}\Leftrightarrow \psi_{1},
$$
and, as a result, they are physically equivalent.

Finally, we will consider the continuum limit, with $a\rightarrow 0$.
The dictionary for taking this limit is
\bq
\sum_{n}&\rightarrow &\frac{1}{a}\int d\sigma,\;\;\;\frac{1}{\sqrt{a}}
\psi_{i}(\sigma_{n},\tau)\rightarrow \psi_{i}(\sigma,\tau),\nonumber\\
\frac{1}{a} \rho(\sigma_{n},\tau)&\rightarrow& \rho(\sigma,\tau)=
\frac{1}{2} \sum_{i}\bar{\psi}_{i}(\sigma,\tau)\psi_{i}(\sigma,\tau).
\eq
For the continuum fermions, we use the same notation as the discrete
ones, but with $\sigma_{n}$ replaced by $\sigma$. They satisfy
anticommutation relations similar to the discretized version, 
 with the Kroenecker delta
replaced by the Dirac delta function. The expressions for $S_{2}$ and
$S_{3}$ (eqs.(17) and (20)) remain unchanged if one replaces the sum over
$\sigma_{n}$ by integration over $\sigma$. In particular, there is
no explicit dependence on $a$. The state corresponding to a solid line
at $\sigma=\sigma'$ is now represented by
$$
|\sigma'\rangle=\bar{\psi}_{1}(\sigma',\tau)\bar{\psi}_{2}(\sigma',
\tau)|0\rangle,
$$
and
\be
\rho(\sigma,\tau)|\sigma'\rangle=\delta(\sigma-\sigma') |\sigma'\rangle.
\ee
 In what follows, although
we will mostly work with the continuum fermions, from time to time
we will also need the world sheet with discretized $\sigma$
 to have well defined expressions.

The total action is the sum of various pieces given by eqs.(13),(17) and
(20):
\be
S=S_{0} + S_{1} + S_{2} + S_{3}.
\ee
As it stands, this action suffers from a serious problem: It is not
well defined. We will discuss this problem and present its solution
in the next section.

We close with a discussion of the scaling properties of the fermions.
 Invariance of the
canonical algebra, or of the free action for the continuum fermions requires
the transformation law
\be
\psi(\sigma,\tau)\rightarrow \sqrt{u}\,\psi(u\sigma,u\tau),\;\;\;
\bar{\psi}(\sigma,\tau)\rightarrow \sqrt{u}\,\bar{\psi}(u\sigma,
u\tau),
\ee
and as result
\be
\rho(\sigma,\tau)\rightarrow u\, \rho(u\sigma,u\tau),\;\;\;
\bar{\rho}(\sigma,\tau)\rightarrow u\,\bar{\rho}(u\sigma,u\tau).
\ee
We also note that, from its definition $N a=p^{+}$, it follows that
the cutoff parameter $a$ must scale as $p^{+}$:
\be
a\rightarrow a/u.
\ee
It is now easy to check that, with the possible exception of the
interaction term $S_{3}$, all the terms of the fermionic action are
scaling invariant. The interaction term violates scaling,
  unless the coupling constant $g$ is allowed to
transform. Of course, the original field theory coupling constant
is a Lorentz scalar and cannot transform. However, the coupling
constant on the world sheet is closely related to but not identical 
to the field theory coupling constant, and it need not be a Lorentz
scalar. In fact, Lorentz invariance
requires that
$$
g\rightarrow u\, g
$$
under scaling. The simplest way to secure this is to use $p^{+}$,
the only other physical parameter at our disposal and make the
replacement
\be
g\rightarrow \frac{g}{p^{+}}.
\ee
$p^{+}$ takes care of the scaling (see eq.(6)), and the newly defined
$g$ is a true scalar.

\vskip 9pt

\noindent{\bf 5. The Gauge Fixed Worldsheet Action}

\vskip 9pt

Consider eq.(16) for $S_{1}$: Since $\rho$ vanishes everywhere except
on the solid lines (the boundary), the integrand is independent of
the Lagrange multipliers ${\bf y}$, $\bar{{\bf b}}$, $\bar{{\bf c}}$
 and ${\bf z}$
in the bulk of the world sheet. As a result, the functional integration
over the bosonic variables is divergent\footnote{I thank Charles Thorn
for stressing this point, although at the time I did not think it was
important.  }
 and the corresponding integration
over the fermionic variables vanishes. The result is an ill defined
 expression of the type infinity times zero. This is similar to what happens in
gauge theories before gauge fixing. In fact, $S_{1}$ is invariant under
the  gauge transformation
\bq
{\bf y}&\rightarrow& {\bf y}+ \bar{\rho}\,{\bf y}_{0},\;\;\;
\bar{{\bf b}}\rightarrow \bar{{\bf b}}+\bar{\rho}\,\bar{{\bf b}}_{0},
 \nonumber\\
{\bf z}&\rightarrow&{\bf z}+\bar{\rho}\,{\bf z}_{0},\;\;\;
\bar{{\bf c}}\rightarrow \bar{{\bf c}}+\bar{\rho}\,\bar{{\bf c}}_{0},
\eq
where ${\bf y}_{0}$, ${\bf z}_{0}$, $\bar{{\bf b}}_{0}$ and $\bar{{
\bf c}}_{0}$ are arbitrary functions and $\bar{\rho}$ is defined by
$$
\bar{\rho}=1-\rho,
$$
for discretized $\sigma$ and
\be
\bar{\rho}=\frac{1}{a}-\rho=\frac{1}{2} \sum_{i}\psi_{i}\bar{\psi}_{i}
\ee
for continuous $\sigma$, where $\rho$ is given by (19) in the first
case and by (21) in the second case.
 In either case, since
\be
\rho\,\bar{\rho}=0,
\ee
invarinace of $S_{1}$ under the transformations (28) follows. It is
this gauge invariance that is responsible for the singular behaviour of
the functional integrals over the Lagrange multipliers mentioned
above.

The cure for this problem is to gauge fix the action, but only in the bulk
of the world sheet (on the dotted lines),
 where $\rho=0$. We will also demand the gauge fixing 
term to be supersymmetric, since our philosophy is to keep intact the
SUSY in the bulk and violate it only on the boundaries. We therefore
promote the Lagrange multipliers into a supersymmetric multiplet
\be
{\bf Y}={\bf z}+\theta_{1}\,\bar{{\bf c}}+\theta_{2}\,\bar{{\bf b}}+
\theta_{1}\theta_{2}\,{\bf y},
\ee
and write the gauge fixing part of the action as
\bq
S_{4}&=&\int_{0}^{p^{+}} d\sigma \int d\tau\int d^{2}\theta
\,\frac{1}{2}\alpha\, \bar{\rho}\,{\bf Y}\cdot {\bf Q}\nonumber\\
&=&\int_{0}^{p^{+}} d\sigma \int d\tau\,\alpha\,\bar{\rho}
\left({\bf y}\cdot{\bf z}+
\bar{{\bf b}}\cdot\bar{{\bf c}}\right),
\eq
where $\alpha$ is a gauge fixing parameter. This is the simplest gauge
fixing term which is supersymmetric and which vanishes on the boundaries
(solid lines)
where $\bar{\rho}=0$ and no gauge fixing is needed.
 The previously singular functional integral over
the bulk of the world sheet where $\rho=0$ is now equal to unity:
\be
\int_{\rho=0}\mathcal{D}{\bf y}\,\mathcal{D}{\bf z}\,\mathcal{D}
\bar{{\bf b}}\,\mathcal{D}\bar{{\bf c}}\,\exp(i S_{4})=1.
\ee
We note that the $\alpha$ dependent integration measures of the
bosonic and fermionic functional integrals cancel. We believe that
the absence of gauge fixing on the continuum world sheet caused
some of the problems encountered in the earlier work.

There is one more technical issue which we have to address here.
There is some arbitrariness
in the equation for (16) for $S_{1}$; it could be replaced by the 
following more general expression:
\be
S_{1}\rightarrow\int_{0}^{p^{+}} d\sigma \int d\tau\,\rho
\left(\beta_{1}\,{\bf y}\cdot\dot{{\bf q}}+\beta_{2}\,
\bar{{\bf b}}\cdot{\bf b}+\beta_{3}\,\bar{{\bf c}}\cdot
{\bf c}+\beta_{4}\,{\bf z}\cdot{\bf r}\right).
\ee
Here, $\beta_{1,2,3,4}$ are arbitrary constants. They
can be eliminated by
 absorbing them into the definition of the Lagrange multipliers.
 So long as one is dealing with the exact expression for
 the action, the introduction of the $\beta$'s changes nothing, and one
could just as well set them all equal to one, as in eq.(16). However,
if an approximation scheme is used, the results may well depend
on these constants. These remarks also apply to the gauge fixing
parameter $\alpha$: The exact theory is independent of this constant
but an approximate calculation may introduce some dependence. In the
next section, when we carry out a mean field calculation, we will
be able to see to what extend our results are sensitive to the
choice of these constants.

Finally, we have to make sure that the gauge fixing term $S_{4}$
 and also $S_{1}$ are scale invariant.  
 This is indeed the case for $S_{4}$ if under scaling
\bq
{\bf y}(\sigma,\tau)&\rightarrow& {\bf y}(u\sigma,u\tau),\;\;\;
\bar{{\bf b}}(\sigma,\tau)\rightarrow \sqrt{u}\,\bar{{\bf b}}
(u\sigma,u\tau),\nonumber\\
\bar{{\bf c}}(\sigma,\tau)&\rightarrow& \sqrt{u}\,\bar{{\bf c}}
(u\sigma,u\tau),\;\;\;{\bf z}(\sigma,\tau)\rightarrow u\,
{\bf z}(u\sigma,u\tau),
\eq
and eq.(25) is taken into consideration. Now that we know the
scaling properties of all the fields, we can check 
$S_{1}$ (eq.(16)). The first three terms
are indeed invariant as they stand, but in the last term,
we have to let
$$
\beta_{4}\rightarrow p^{+}\,\beta_{4}.
$$

\vskip 9pt

\noindent{\bf 6. The Mean Field Approximation}

\vskip 9pt

In the last section, the final form of the action which
is supersymmetric, gauge fixed and scale invariant was
worked out. Since the mean field approximation will be
based on it,
we start by writing it down in full:
\bq
S&=& \sum_{n=0}^{n=4} S_{n}\nonumber\\
&=&\int_{0}^{p^{+}} d\sigma \int d\tau\Big(-\frac{1}{2}(
{\bf q}'^{2}+{\bf b}\cdot{\bf b}'+{\bf c}\cdot{\bf c}'
+{\bf r}^{2})+\rho (\beta_{1}\,{\bf y}\cdot\dot{{\bf q}}+
\beta_{2}\,\bar{{\bf b}}\cdot{\bf b}\nonumber\\
&+&\beta_{3}\,\bar{{\bf c}}\cdot{\bf c}
+\beta_{4}\,p^{+}\,{\bf z}\cdot{\bf r})
+\alpha \bar{\rho}\,({\bf y}\cdot{\bf z}+\bar{{\bf b}}\cdot\bar{{\bf c}})
+i \bar{\psi}\dot{\psi}+\frac{g}{p^{+}}(\bar{\psi}_{1}\bar{\psi}_{2}+
\psi_{2}\psi_{1})\Big),\nonumber\\
& &
\eq   
where $\rho$ and $\bar{\rho}$ are given by (21) and (29).

The mean field approximation was developed and applied to the
world sheet action in the earlier work [2,3,4]. Here, we need the
simplest version of it used in [2]. We notice that eq.(36) represents
a vector model, which can be solved in the large $D$ limit [16].
 The standard
approach is  to replace scalar products of various vector
fields, such as the bilinear terms in the expression for $S_{0}$
(eq.(13)), by their vacuum expectation values, which are then treated as
classical c-number fields. The functional integral over the
remaining fields is carried out exactly, and the resulting
effective action is minimized with respect to the 
classical fields. This method is known to be equivalent to
a large $N$ ( in this case, $D$ replaces $N$) saddle point
calculation [16]. Instead of the approach sketched above,
we find it much simpler to replace the bilinears in the
fermions, $\rho$ and $\bar{\rho}$ by their classical expectation
values. In the Appendix, we show that this is completely equivalent to
replacing the bilinears in vector fields by their expectation values.

The problem is considerably simplified by starting with the setup
where the total transverse momentum ${\bf p}$ carried by the whole
graph is zero:
\be
{\bf p}=\int_{0}^{p^{+}} d\sigma\,{\bf q}'=0.
\ee
This configuration can always be reached 
by a suitable Lorentz transformation. It allows us
to impose the periodic boundary conditions
\be
{\bf q}(\sigma=0,\tau)={\bf q}(\sigma=p^{+},\tau).
\ee
This setup is translationally invariant in both the $\sigma$ and
the $\tau$ directions, and we shall see that translation invariance
will play an important role in simplifying the mean field calculation.

We start by explicitly introducing the composite field $\rho$
by adding a new term $\Delta S$ to the action:
\bq
S&\rightarrow& S+\Delta S,\nonumber\\
\Delta S&=&\int_{0}^{p^{+}} d\sigma\int d\tau\, \kappa\left(\frac{1}{2}\sum_{i}
\bar{\psi}_{i}\psi_{i}-\rho\right),
\eq
where $\kappa$ is a Lagrange multiplier, and $\bar{\rho}$ is  given
in terms of $\rho$ through eq.(29). In the large $D$ limit, we can treat
$\kappa$ and $\rho$ as classical fields (See the Appendix). In other
words, we make the replacement
\be
\kappa\rightarrow \kappa_{0}=\langle \kappa \rangle,\;\;\;
\rho\rightarrow \rho_{0}=\langle \rho \rangle,\;\;\;
\bar{\rho}\rightarrow \bar{\rho}_{0}= \langle \bar{\rho} \rangle,
\ee
where $\langle\rangle$ denotes the expectation value in the ground 
state of the field in question. Translation
invariance on the world sheet means that both $\kappa_{0}$ and $\rho_{0}$ are
constants independent of the coordinates $\sigma$ and $\rho$. With this
simplication, it is possible to carry out explicitly all the functional
integration over the fields. We first consider the integration over the
fermions; that part of the action involving the fermions is given by
\be
S_{f}=\int_{0}^{p^{+}} d\sigma\int d\tau\left(i\bar{\psi}\dot{\psi}
+\frac{g}{p^{+}}
(\bar{\psi}_{1}\bar{\psi}_{2}+\psi_{2}\psi_{1})+\frac{\kappa_{0}}{2}
\sum_{i}\bar{\psi}_{i}\psi_{i}\right).
\ee
Instead of working with the action, we find it more convenient to
diagonalize the corresponding Hamiltonian. In order to avoid singular
expressions, we first discretize the $\sigma$ coordinate as in eqs.(17)
and (20). There is a complete decoupling of the dynamics in the $\sigma$
direction; as a result, the total Hamiltonian can be written as a sum of
$N$ mutually commuting Hamiltonians:
\be
H=\sum_{n} H_{n},
\ee
where,
\be
H_{n}=- \left(g' (\bar{\psi}_{1}\bar{\psi}_{2}+ \psi_{2}\psi_{1})+
\frac{\kappa_{0}}{2} \sum_{i}\bar{\psi}_{i}\psi_{i}\right)_{\sigma=
\sigma_{n}},
\ee
and
$$
g'=g/p^{+}.
$$

We observe that $H_{n}$ acts on the two states
$$
|0\rangle
$$
corresponding to a dotted line at $\sigma=\sigma_{n}$ and
$$
|n\rangle= \bar{\psi}_{1}(\sigma_{n})\bar{\psi}_{2}(\sigma_{n})
|0\rangle
$$
corresponding to a solid line at the same position as a two by two matrix:
\bq
H_{n}|0\rangle &=& -g' |n\rangle,\nonumber\\
H_{n}|n\rangle &=& -g' |0\rangle -\kappa_{0} |n\rangle.
\eq
The eigenvalues are
\be
E^{\pm}_{f}=-\frac{1}{2} \kappa_{0} \pm\frac{1}{2} \sqrt{\kappa_{0}^{2}+
4 g'^{2}},
\ee
where the square root is defined to be positive.
Since we are  interested in the ground state, we have to pick
the lower energy
$E^{-}_{f}$ at each $\sigma=\sigma_{n}$ and add up to get the
total fermionic energy:
\be
E_{f}= N E_{f}^{-}=\frac{p^{+}}{a} E_{f}^{-}.
\ee

Next, we will carry out the functional integrations over the vector
fields in eq.(36), setting
$$
\rho\rightarrow \rho_{0},\;\;\;\bar{\rho}\rightarrow \bar{\rho}_{0}
=\frac{1}{a}- \rho_{0}.
$$
 We first split the action into $S_{m}$, the matter
action, and $S_{g}$, the ghost action:
\bq
S_{m} &=& \int_{0}^{p^{+}} d\sigma \int d\tau\left(-\frac{1}{2}
({\bf q}'^{2}+{\bf r}^{2})+ \rho_{0} (\beta_{1}\,{\bf y}\cdot \dot{{
\bf q}} +\beta_{4}\,p^{+}\,{\bf z}\cdot {\bf r})+ \alpha \bar{\rho}_{0}
{\bf y}\cdot {\bf z}\right),\nonumber\\
S_{g} &=& \int_{0}^{p^{+}} d\sigma \int d\tau \left(-\frac{1}{2}
({\bf b}\cdot {\bf b}'+ {\bf c}\cdot {\bf c}')+ \rho_{0} (\beta_{2}\,
\bar{{\bf b}}\cdot {\bf b}+ \beta_{3}\, \bar{{\bf c}}\cdot
{\bf c})+\alpha \bar{\rho}_{0}\,\bar{{\bf b}}\cdot \bar{{\bf c}}\right).\nonumber\\
& &
\eq
The integrations over ${\bf y}$, ${\bf r}$, $\bar{{\bf b}}$ and
$\bar{{\bf c}}$ are Gaussian and they can be done trivially. The
(singular) Jacobians coming from integrations over the matter and
the ghost fields cancel, with the result,
\bq
S_{m} &\rightarrow & \int_{0}^{p^{+}} d\sigma \int d\tau \left(
-\frac{1}{2} {\bf q}'^{2} + \frac{1}{2}\Big(\frac{\beta_{1} \beta_{4}
p^{+} \rho_{0}^{2}}{\alpha \bar{\rho}_{0}}\Big)^{2} \dot{{\bf q}}^{2}\right),
\nonumber\\
S_{g} &\rightarrow& \int_{0}^{p^{+}} d\sigma \int d\tau \left(
-\frac{1}{2}{\bf b}\cdot {\bf b}' -\frac{1}{2} {\bf c}\cdot
{\bf c}'+ \frac{\beta_{2} \beta_{3} \rho_{0}^{2}}{\alpha \bar{\rho}_{0}}\,
{\bf c}\cdot {\bf b}\right).\nonumber\\
& &
\eq

We note that $S_{m}$ is the(Minkowski) world sheet action for a string, with the
slope parameter $\alpha'$, where
\be
 4 \alpha'^{2}= \left(\frac{\beta_{1}\beta_{4} p^{+} \rho_{0}^{2}}{\alpha \bar
{\rho}_{0}}\right)^{2}.
\ee
It may seem like string formation is almost automatic;
 however, the string picture breaks down if the slope
is  zero , which happens for $\rho_{0}=0$. The parameter $\rho_{0}$
is therefore the order parameter that distinguishes between the
stringy and the perturbative phases of the same field theory.
 Roughly speaking, since $\rho_{0}$ measures
the fraction of the world sheet area occupied by the solid lines,
each graph in perturbation theory corresponds to $\rho_{0}=0$.
This is because any finite number of solid lines, being one
dimensional, have vanishing area, and as to be expected, perturbative
field theory is then the zero slope limit of the string theory. A non-
zero slope requires $\rho_{0}\neq 0$, which means that the solid
lines condense to occupy a finite fraction of the area of the world
sheet. Therefore, $\rho_{0}$ serves
 as an order parameter which
distinguishes between two phases: $\rho_{0}=0$ in the perturbative
field theory phase and $\rho_{0}\neq 0$ in the stringy phase.
 In the next section, we will find that $\rho_{0}\neq 0$
 in the mean field approximation.

Since $S_{m,g}$ have quadratic dependence on the fields, the functional
integrals can be carried out. Defining
$$
\int \exp(i S_{m,g})=\exp(i S^{e}_{m,g}),
$$
we have, after Euclidean rotation in $\tau$,
\bq
S_{m}^{e}&=& -\frac{1}{2} D\,Tr\ln\left(-\partial_{\sigma}^{2} - A^{2}
\partial_{\tau}^{2}\right),\nonumber\\
S_{g}^{e}&=& D\,Tr\ln\left(-\partial_{\sigma}^{2}- B^{2} \partial^{2}
_{\tau}\right),
\eq
where,
$$
A=\frac{\beta_{1}\beta_{4} p^{+}\rho_{0}^{2}}{\alpha \bar{\rho}_{0}},
\;\;\;B=-i \frac{\beta_{2}\beta_{3}\rho_{0}^{2}}{\alpha \bar{\rho}_{0}}.
$$
Rewriting the $Tr\ln$'s in terms of momenta $k_{0}$ and $k_{1}$ conjugate
to $\tau$ and $\sigma$ gives
\be
S^{e}= S_{m}^{e}+S_{g}^{e}=D\,\frac{\tau_{f}-\tau_{i}}{4\pi}\int d k_{0}
\sum_{n}\ln\left(\frac{(k_{1}^{n})^{2}+ B^{2}}{(k_{1}^{n})^{2}+ A^{2}
k_{0}^{2}}\right),
\ee
with
$$
k_{1}^{n}=\frac{2\pi n}{p^{+}}.
$$
This is only a formal result, since the integration over $k_{0}$ and
 summation over $n$ lead to a quadratic divergence, and before one
can make sense of it, it must be regulated. In fact, we are precisely
interested in the coefficient of this quadratic divergence, which, after
it is regulated by a cutoff, will be the dominant term in the answer.
 Since the answer is somewhat sensitive to  the cutoff procedure,
  we will try to explain the motivation for  the regulator we use. First,
 we make use of a simplification:  Only the leading cutoff
dependent part of the answer is of interest and  this dependence
 is sensitive
only to the large $k_{1}^{n}$ and large $k_{0}$ behaviour of the
integrand in the above equation.
Therefore, we can safely replace the summation over $n$ by
the integration over the continuous variable $k_{1}$. By the same
token, we can set $B=0$ in the integrand:
\be
S^{e}\rightarrow (\tau_{f}-\tau_{i}) \frac{D\,p^{+}}{8\pi^{2}} 
\int d k_{0} \int d k_{1}\ln\left(\frac{k_{1}^{2}}
{k_{1}^{2}+ A^{2} k_{0}^{2}}\right).
\ee

In eq.(51), the contributions of the
matter and ghost sectors were combined into a single $\log$.
This was intentional: We are going to regulate the combined
contributions of the two sectors, rather then regulate them
seperately. To see why, we observe that\\
a) The combined term is less singular then each
term treated seperately. In fact, for a fixed
$k_{0}$, the integral over $k_{1}$ converges, so that we
need to regulate only the integral over $k_{0}$. This is no
accident: it can be traced back to the cancellation of the
singular determinants between the matter and the ghost sectors.
Regulating each term seperately could spoil this cancellation.\\
b) We saw in section 2 that scale invariance on the world sheet
is necessary for Lorentz invariance in the target space. Therefore,
the regulated expression for $S^{e}$ should be scale invariant.
Under scaling, $k_{0}$ and $k_{1}$ transform as
$$
k_{0}\rightarrow u\,k_{0},\;\;\;k_{1}\rightarrow u\,k_{1},
$$
and from eq.(50), it is easy to check that $A$ is scale invariant.
It then follows that the integrand in eq.(52) is scale invariant.
This is of course related to the cancellation discusssed above,
if we recall from section 3 that the cancellation of determinants
and scale invariance are intimately connected. Again, it was
essential to combine the matter and the ghost terms to arrive at
a scale invariant integrand.

The scale invariance of the integrand in eq.(52) is necessary but
not sufficient for the scale invariance of $S^{e}$; one also
needs a regulator that respects scaling. We have seen that only
the $k_{0}$ integration has to be regulated, which we regulate
by introducing
a second cutoff (in addition to $a$) in the $\tau$ direction.
Again the precise form of the regulator is not important, so
long as it respects scaling. As
a simple example, we will consider a sharp cutoff in $k_{0}$:
\be
S^{e}\rightarrow (\tau_{f}-\tau_{i})\frac{D\,p^{+}}{8\pi^{2}}
\int_{-\lambda/p^{+}}^{\lambda/p^{+}} dk_{0} \int dk_{1}
\ln\left(\frac{k_{1}^{2}}{k_{1}^{2}+ A^{2} k_{0}^{2}}\right),
\ee
where $\lambda$ is the cutoff parameter. The factor of $p^{+}$
is inserted so that the limits of $k_{0}$ integration are
invariant under scaling. By a change of variables, we can
evaluate this integral as
\be
S^{e}=- \frac{D}{8\pi^{2}}\frac{\tau_{f}-\tau_{i}}{p^{+}}
A\,\Lambda,
\ee
where,
\be
\Lambda=\int_{-\lambda}^{\lambda} dk_{0} \int dk_{1}
\ln\left(\frac{k_{1}^{2}+ k_{0}^{2}}{k_{1}^{2}}\right).
\ee

We note that\\
a) $\Lambda$ is a positive cutoff dependent constant. Since
this is the only cutoff dependent parameter in the result,
we may as well replace the original cutoff parameter
$\lambda$ by $\Lambda$.\\
b) The simple linear dependence of $S^{e}$ on $A$ is
going to be important in the following development. 
 This dependence is a fairly robust result:
It follows from the change of variable
$$
k_{1}\rightarrow A\,k_{1},
$$
independent of the details of how the integral over
$k_{0}$ is regulated.\\
c) The dependence on $p^{+}$ is required by scale invariance,
again independent of the form of the regulator.\\
The preceding discussion leads to the conclusion that
$S^{e}$ has the unique form given by eq.(54), provided that
we combine the matter and ghost determinants before
regularizing and we use a regulator that respects scale
invariance.

Finally, we would like to comment on the factor of $i$ in 
the definition of $B$ in eq.(54). If the product $\beta_{2}
\beta_{3}$ is real, $B^{2}$ will be negative, and from
eq.(), $S^{e}$ will be complex. This is, of course, a 
signal for instability. This does not concern us here: Since  
we are interested only in the leading cutoff dependence,
 we use eq.(51), where the $B^{2}$ term is absent. However,
even for the non-leading part of $S_{e}$, it is possible
to avoid this problem by taking the product $\beta_{2}
\beta_{3}$ to be pure imaginary. We recall that the constants
$\beta_{2}$ and $\beta_{3}$ appeared in front of the Lagrange 
multipliers that set the fields ${\bar b}$ and ${\bar c}$
equal to zero. If these fields were bosonic, complex values for
these constants would not be permissible, but for fermionic
fields, complex coefficients are allowed. Nevertheless, this may
still show up as an instability in some non-leading order
in the large $D$ limit.

\vskip 9pt

\noindent{\bf 7. String Formation}

\vskip 9pt

We can now put together various terms derived in the last
section and write down the full effective action $S^{eff}$:
\bq
S^{eff}&=& S^{e}- (\tau_{f}-\tau_{i})(E_{f}+ p^{+}\,
\kappa_{0} \rho_{0})=\frac{\tau_{f}-\tau_{i}}
{p^{+}} \tilde{S},\nonumber\\
\tilde{S}&=&-\frac{D}{8\pi^{2}}\frac{\beta_{1}\beta_{4} p^{+}
\rho_{0}^{2}}{\alpha \bar{\rho}_{0}} \Lambda+ \frac{(p^{+})^{2}}
{2 a}\left(\kappa_{0}+\sqrt{\kappa_{0}^{2}+4 g'^{2}}\right)
-(p^{+})^{2} \kappa_{0} \rho_{0}.\nonumber\\
& &
\eq
Now let us go back to the question posed in Section 5: How does
the effective action depend on the arbitrary constants
$\beta_{1,2,3,4}$ that appear in eq.(36) and the gauge fixing
parameter $\alpha$? The answer is that, at least within
the present approximation, it does not depend on $\beta_{2}$
and $\beta_{3}$ at all. Also, the dependence on $\alpha$ and
$\beta_{1}$ and $\beta_{4}$ is rather trivial. Since the cutoff
parameter $\Lambda$ is arbitrary to begin with, by redefining
it through
$$
\Lambda\rightarrow \frac{\beta_{1}\beta_{4}}{\alpha}\Lambda,
$$
one can eliminate all the reference to these parameters.

Next we will search for the saddle point of the effective
action in the variables $\rho_{0}$ and $\kappa_{0}$,
or equivalently in the variables $x$ and $y$, and
see whether this corresponds to the minimum value of the
ground state 
energy. To start with, the expression for $\tilde{S}$ can
be simplified considerably by a series of redefinitions:
\bq
 \tilde{\Lambda}&=&\frac{\beta_{1}\beta_{4}}{8\pi^{2}\,\alpha}
 \Lambda,\;\;\rho_{0}=\frac{x}{a},\nonumber\\
\bar{\rho_{0}}&=&\frac{1-x}{a},\;\;\kappa_{0}=2 \frac{D \tilde{
\Lambda}}{p^{+}}\,y,\;\;g'=\frac{g}{p^{+}}=\frac{D \tilde{\Lambda}}
{p^{+}} \tilde{g}.
\eq
It is also convenient to define a scale invariant cutoff
parameter $a'$ by
$$
a'=a/p^{+}.
$$
In terms of these new variables, $\tilde{S}$ is given by
\be
\tilde{S}=D\,\frac{\tilde{\Lambda}}{a'} F(x, y),
\ee
where
\be
F(x, y)= -\frac{x^{2}}
{1-x} -2 x y +y +\sqrt{y^{2}+ \tilde{g}^{2}},
\ee
and $y$ ranges from $-\infty$ to $+\infty$, whereas
$x$ is limited to the interval $0\leq x\leq 1$.

A number of features of this expression are worth noting:\\
a) There is a factor of $D$ multiplying the whole expression.
Therefore, in the large $D$ limit, the dominant contribution
comes from the saddle point.\\
b) Every variable in this expression is scale invariant.\\
c) The factor of $D$ appearing in the definition
of $\tilde{g}$ is the standard ``large N'' factor [8]
needed to have the correct limit.\\
d) Although we have so far introduced two independent
cutoff parameters $\Lambda$ and $a$, or equivalently,
$\tilde{\Lambda}$ and $a'$, only the combination
$\tilde{\Lambda}/a'$ appears in the expression for
$\tilde{S}$.

The integral to be evaluated in the large $D$ limit is
\bq
Z&=&\int_{-\infty}^{+\infty} dy \int_{0}^{1} dx\,
\exp(i S^{eff}),\nonumber\\
S^{eff}&=&\gamma\, F(x,y),
\eq
and the constant $\gamma$ is defined by
$$
\gamma= D\,\frac{\tilde{\Lambda}}{a'}\frac{\tau_{f}
-\tau_{i}}{p^{+}}.
$$
First, let us consider the integration over $y$ for
a fixed value of $x$ in the interval (0,1). This integral
can be evaluated exactly in terms of a Bessel function,
but since we need only the large $D$ result, we will instead
use the saddle point approximation.
 The saddle
point  $y_{s}$ is at
\be
\partial_{y}F(x,y)=0 \rightarrow y_{s}=\frac{2 x-1}
{2\sqrt{x -x^{2}}}\, \tilde{g}.
\ee
In this equation, both $\tilde{g}$ and the square root
are defined to be positive. The contour of integration
in the complex $y$ plane can be distorted into the curve
whose equation is
\be
Re\left(-2 x y+ y +\sqrt{y^{2}+ \tilde{g}^{2}}\right)
=2 \tilde{g}\sqrt{x -x^{2}}.
\ee
This curve passes through the saddle point, and with the
above choice of the branch of the square root, as
$$
Im(y)\rightarrow \pm\infty,
$$
on the curve,
$$
Im\left(F(x,y)\right)\rightarrow +\infty,
$$  
and therefore the integral
$$
\int_{y=curve} dy\,\exp\left(i\gamma F(x,y)\right)
$$
is exponentially convergent. Using this contour of
integration justifies the evaluation of the integral over $y$
by setting $y=y_{s}$ in the integrand.
In Fig.4, the curve defined by eq.(62) in the complex
$y$ plane is pictured for $\tilde{g}=1$ and
$x=1/2$, when the saddle point is at $y_{s}=0$.
\begin{figure}[t]
\centerline{\epsfig{file=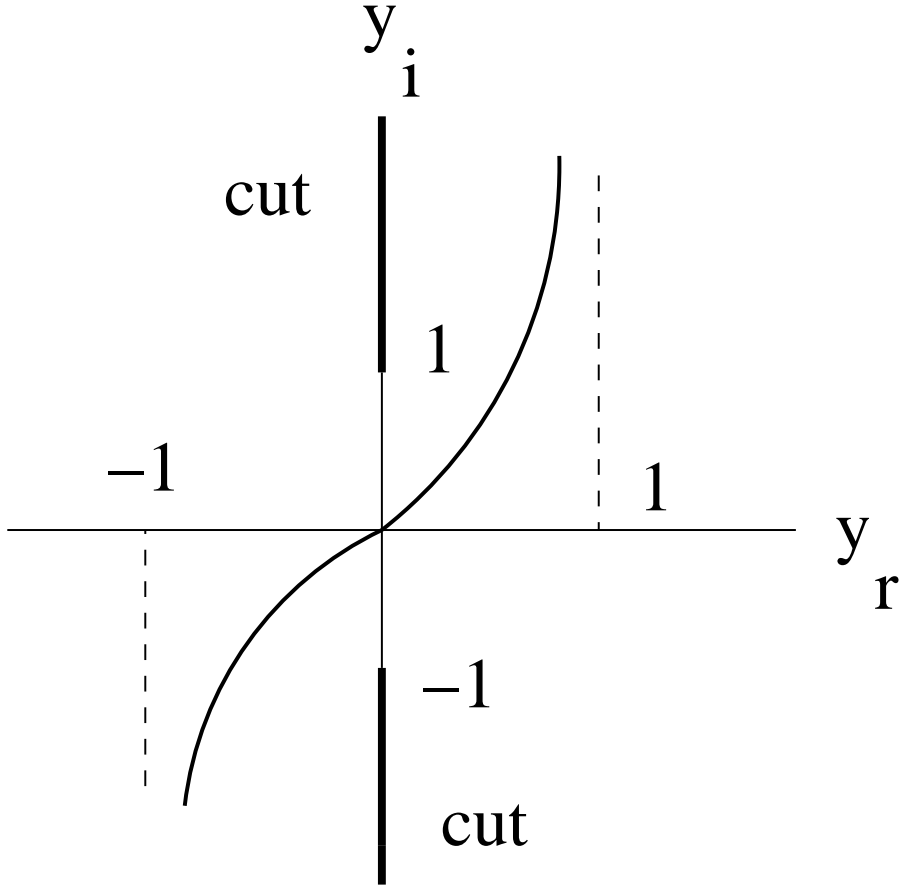, width=12 cm}}
\caption{The Integration Contour in the Complex y Plane
for $\tilde{g}=1$}
\end{figure}
 The branch
cuts of the square root run from $i\tilde{g}$ to $+i\infty$
and from $-i\tilde{g}$ to $-i\infty$ and the contour
asymptotes the vertical lines $Re(y)=\pm\tilde{g}$.

After the saddle point evaluation of the integral over $y$,
we are left with the integral over $x$:
\be
Z\rightarrow \int_{0}^{1} dx \exp\left(- i \gamma\, f(x)\right),
\ee
where
\be
f(x) = \frac{x^{2}}{1 -x} - 2 \tilde{g} \sqrt{x- x^{2}}.
\ee
The function $f(x)$ is pictured in Fig.5 for $\tilde{g}=20$.
\begin{figure}[t]
\centerline{\epsfig{file=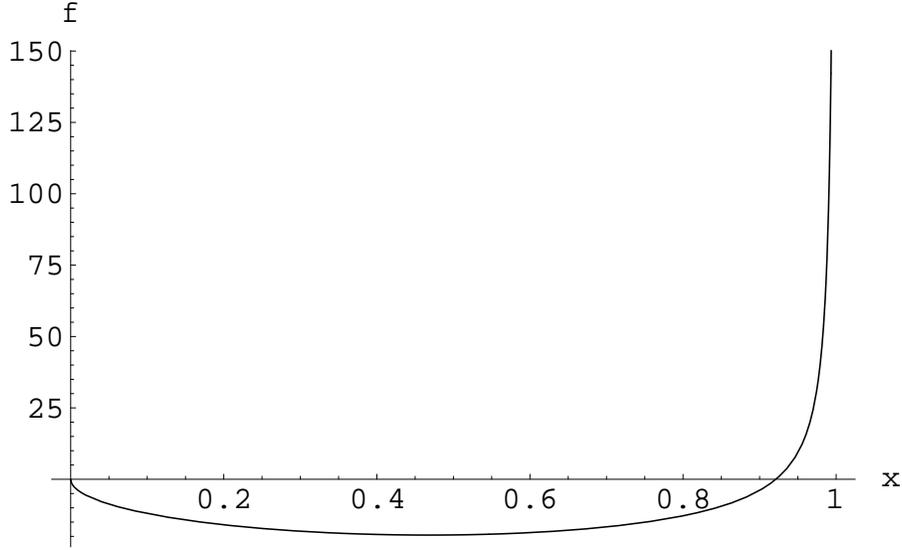,width=12cm}}
\caption{The Function f(x) for $\tilde{g}=20$}
\end{figure}
It has a single minimum in the interval $0\leq x\leq 1$ at 
$x=x_{m}$, which satisfies
\be
f'(x_{m})=\frac{2 x_{m}- x_{m}^{2}}{(1 -x_{m})^{2}}-\tilde{g}
\frac{1- 2 x_{m}}{\sqrt{x_{m} - x_{m}^{2}}}=0.
\ee
At the minimum, $f(x)$ is negative, corresponding to a
negative minimum energy
$$
E_{0}=D\,\frac{\tilde{\Lambda}}{a'\,p^{+}}\, f(x_{m}),
$$
which is the energy of the ground state in this approximation.
The situation is similar for other positive values of
$\tilde{g}$: There is only a single minimum in the interval $(0,1)$,
corresponding to a negative ground state energy.

We have just seen that the ground state corresponds to a
value of $x$ between 0 and 1. Let us remember that
$$
x=0 \rightarrow \rho_{0}=0
$$
corresponds to the trivial case of a world sheet with all
dotted lines. The opposite limit of
$$
x=1 \rightarrow \bar{\rho}_{0}=0
$$
corresponds to a world sheet with only solid lines. A value
of $x$ in between these two extremes implies an intermediate 
world sheet texture: The solid and dotted lines each occupy
a finite fraction of the area of the world sheet. Recalling
our earlier discussion following eq.(49), we see that indeed 
a condensate of the solid lines has formed.

Let us now see whether a sensible string picture emerges.
In particular, the slope parameter (eq.(49))
\be
\alpha'=\frac{\beta_{1}\beta_{4} p^{+} \rho_{0}^{2}}
{2 \alpha \bar{\rho}_{0}}=\frac{\beta_{1}\beta_{4} x^{2}_{m}}
{2 \alpha (1-x_{m}) a'},
\ee
is the only physical parameter to emerge from the mean field
calculation. The theory should be renormalized by requiring it
to be a finite number
independent of any cutoff. To see how this happens, we first get rid
of the irrelevant constants $\beta_{1,4}$ and $\alpha$ by suitably
redefining the cutoff parameter $a'$:
\be
\alpha'\rightarrow \frac{x^{2}_{m}}{2 (1- x_{m}) a'}.
\ee
In order to have a finite slope in the limit $a'\rightarrow 0$,
 $x_{m}$, and therefore, $\tilde{g}$ should
also go to zero in the same limit.
 Solving (65) in the small $\tilde{g}$ limit, we have
\be
x_{m}\approx \left(\frac{\tilde{g}}{2}\right)^{2/3},
\ee
and
\be
\alpha'\approx \frac{\tilde{g}^{4/3}}{2^{7/3}\,a'}.
\ee
Therefore, in the limit of the cutoff $a'$
tending to zero, the coupling constant $\tilde{g}$ should
be fine tuned so that the ratio
$$
\frac{\tilde{g}^{4/3}}{a'}
$$
stays constant. The theory is renormalized by trading
the cutoff dependent coupling constant $\tilde{g}$ for
the cutoff independent physical parameter $\alpha'$
through eq.(69).

There is one more comparison one can make between
the parameters of the $\phi^{3}$ theory and the
string theory: One could try to relate the string
intercept to the field theory
 mass and the coupling constant. If we identify
the mass square of the lowest lying state on the string
trajectory with the ground state energy corresponding
to $\tilde{S}$ (eq.58), the meanfield approximation
gives a negative cutoff dependent answer. In this article,
we will not consider the question of renormalization of
the ground state energy, which is the same as the 
renormalization of the intercept. There is, however,
the following simple possibility. Instead of starting
with a $\phi^{3}$ theory with zero bare mass, we could have
started with a non-zero cutoff dependent bare mass.
Fine tuning this bare mass, it may be possible
to end up with a finite renormalized ground state energy.
 To carry out this program, however,
 our formalism has to be extended to include
a non-zero bare mass. This we leave to future research.

It may seem surprising that, starting with an unstable
field theory, so far we have not
encountered any sign of instability on the string side.
Of course, as mentioned above, 
we have  not  calculated the renormalized intercept.
In the end, upon calculating this intercept,
just as in the case of the bosonic string,
some of the lowest lying states may turn out to be
tachyonic. Another possibility is that, in the leading
order of the large $D$ approximation, the instability
may not be visible \footnote{I would like to thank
Jeff Greensite and Charles Thorn for this suggestion.}.
For example, it is easy to construct a simple quantum
mechanics problem with $D$ degrees of freedom,
 where there is a metastable state
which decays by tunneling\footnote{I would like to
acknowledge a helpful conversation with Eliezer
Rabinovici on this subject.}. It is usually the case
that tunneling is suppressed in the leading
 large $D$ limit, and the metastable state becomes
stable. One has to go beyond the leading order to
see signs of instability. It is possible that this is
what happens in the model we are studying here.

\vskip 9pt

\noindent{\bf 8. An Additional String Mode}

\vskip 9pt

In this section, we are going to compute a particular
 correction to the leading mean field or large $D$ result
\footnote{See also Appendix B of reference [2].}.
We recall that, in this limit, the composite field $\rho$ can be
replaced by its ground state expection value $\rho_{0}$, but to
go beyond the leading term, one has to expand in powers of the
fluctuations around the mean. Of course, an exact computation
of the full expansion is impractical; however, as in section
6, we look for the dominant contribution in the limit of
large cutoff. In this case, this is a logarithmically
divergent term which dominates the rest of the terms, which
are finite. To isolate this contribution, we split
the composite field $\rho$ into the constant mean value $\rho_{0}$
(see eq.(70)), and a fluctuating part $\chi$. Along with a power series
expansion in $\chi$, we treat $\chi$ as a slowly varying function,
and we expand it around a fixed point in powers of the coordinates
$\sigma$ and $\tau$. This latter expansion, which is sometimes
called the derivative expansion,
 is very useful in
isolating divergent terms in the perturbation series and it is widely
used in literature. The point is that increasing order in this
 expansion goes with increasing number of derivatives on $\chi$,
and by dimensional arguments, this results in greater convergence.
The divergent terms therefore appear only in the lowest orders of
the derivative
 expansion and they are easy to identify. We shall see that
 the leading divergence is quadratic, but this is already included
in the calculation done in section 6 with a constant $\rho_{0}$.
The next leading divergence is logarithmic, which is the
contribution we are going to calculate. The rest of the terms
in the expansion are finite. The logarithmic term has a special
physical significance: it provides a kinetic energy term for
$\chi$ in the action and so it promotes $\chi$ into
a new propagating degree of freedom.  This
phenomenon should be familiar from other two dimensional
models, such as the $CP(N)$ model [17] or the Gross-Neveu model [18].  
In contrast, the finite terms in the expansion are non-local
and they do not seem to have any special physical significance.
  
 Our starting point is eq.(48) for $S_{m}$, but with $\rho_{0}$
replaced by $\rho$, since we are considering fluctuations
of $\rho$ around the mean value $\rho_{0}$. We define
\be
\left(\frac{\beta_{1}\beta_{4}\,p^{+} \rho^{2}}{\alpha \bar{\rho}}\right)^{2}
=A^{2}(1+  \chi),
\ee
where $A$ is defined by eq.(50) and,
\be
\chi=\left(\frac{\bar{\rho}_{0} \rho^{2}}{\bar{\rho} \rho_{0}^{2}}
\right)^{2}-1,
\ee
is the fluctuating field. 
 Doing the functional integral over ${\bf q}$ gives
the following contribution to the action:
\be
S'=\frac{i}{2}\,D\, Tr\ln\left(\partial_{\sigma}^{2}
-  A^{2} \partial_{\tau}(1+\chi)\partial_{\tau}\right).
\ee
We are going to examine in detail only terms up to second
order in $\chi$; it will then be easy to figure out the
contribution of the higher order terms. We therefore
expand $S'$ up to second order: 
\bq
S'&=& S'_{0}+ S'_{1} + S'_{2}+\ldots,\nonumber\\
S'_{0}&=&\frac{i}{2}\,D\,Tr\ln\left(\partial_{\sigma}^{2}
- A^{2}\partial_{\tau}^{2}\right),\nonumber\\
S'_{1}&=&-\frac{i}{2}\,D\,A^{2}\,Tr\left((\partial_{\sigma}^{2}
- A^{2} \partial_{\tau}^{2})^{-1} \,\partial_{\tau}\chi\partial_
{\tau}\right),\nonumber\\
S'_{2}&=& \frac{i}{4}\,D\,A^{4}\,Tr\left((\partial_{\sigma}^{2}
- A^{2}\partial_{\tau}^{2})^{-1}\,\partial_{\tau}\chi\partial_{\tau}
\,(\partial_{\sigma}^{2}- A^{2} \partial_{\tau}^{2})^{-1}
\partial_{\tau}\chi\partial_{\tau}\right).
\eq

The zeroth order term $S'_{0}$ was called $S^{e}_{m}$ in section
6 and it was already calculated there (eq.(50)).
 The first order term $S'_{1}$ is
represented by the graph in Fig.6, where the external line
in this graph carries zero momentum.
\begin{figure}[t]
\centerline{\epsfig{file=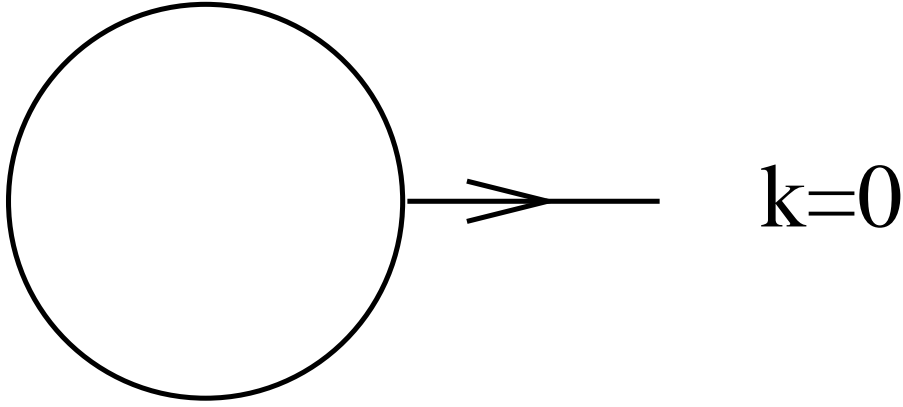,width=12cm}}
\caption{First Order Tadpole Graph}
\end{figure}
 Its contribution is
given by
$$
S'_{1}= c_{1}\int d\sigma \int d\tau\, \chi(\sigma,\tau),
$$
where $c_{1}$ is a quadratically divergent constant. In addition,
there other contributions from higher order terms, represented
by graphs of the form given in Fig.7,
\begin{figure}[h]
\centerline{\epsfig{file=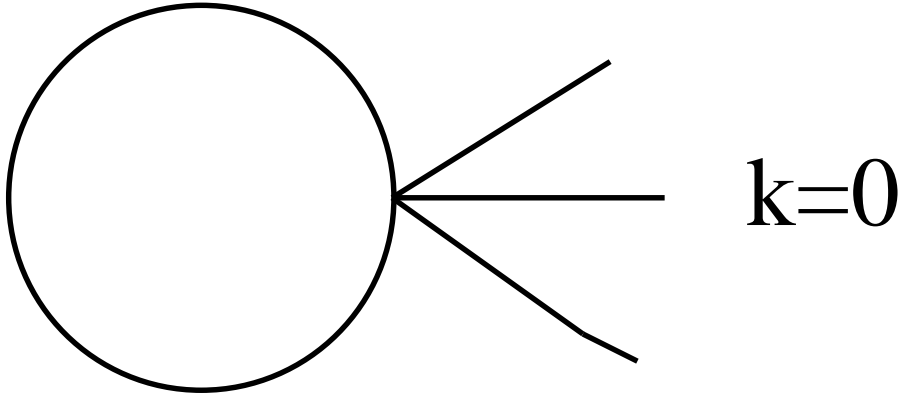,width=12cm}}
\caption{Higher Order Tadpole Graph}
\end{figure}
 which generate the following
series in $S'$:
\be
S'\approx \int d\sigma \int d\tau \sum_{n=1}^{\infty}
c_{n}\left(\chi(\sigma,\tau)\right)^{n}.
\ee
This can be thought of as a potential for $\chi$, which,
when minimized, determines the expectation value of $\chi$.
However, all this does is to shift the value of $\rho_{0}$,
which we have already calculated in the previous section.
This ambiguity is due to the intrinsic arbirariness in
the split made in eq.(70): Only the particular combination of
$\rho_{0}$ and $\chi$ that appears on the right hand side
of (70) is well defined: Shifting $\rho_{0}$ and the
expectation value of $\chi$ while keeping the right hand side
of (70) constant will not change anything.
We can resolve this ambiguity by setting the expectation
value of $\chi$ equal to zero. In that case, there is no
shift in the value of $\rho_{0}$, and
 we can   drop the terms
given in eq.(74). We shall do so in what follows.

Let us now focus on $S'_{2}$, the term quadratic in
$\chi$, which is represented by the graph in Fig.8.
\begin{figure}[b]
\centerline{\epsfig{file=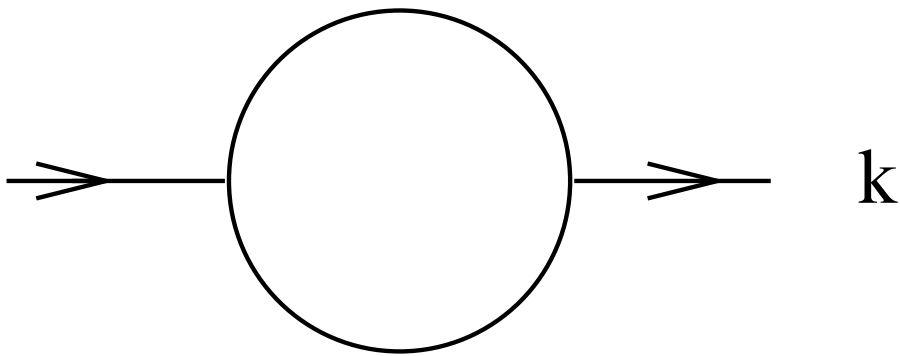,width=12cm}}
\caption{One Loop Contribution to the Propagator}
\end{figure}
In momentum space, we can set
\be
S'_{2}=-\frac{i}{4}\,D\,A^{4}\frac{p^{+}}{2\pi^{3}}
\int d^{2} k' I(k'_{0},k'_{1})\tilde{\chi}(k'_{0},k'_{1})
\tilde{\chi}(-k'_{0},-k'_{1}),
\ee
where $\tilde{\chi}$ is the Fourier transform of $\chi$,
and
\be
I(k_{0},k_{1})=\int d^{2} k \frac{(4 k_{0}^{2}-( k'_{0})^{2})^{2}}
{\Big((2 k_{1}+ k'_{1})^{2} -A^{2}(2 k_{0}+ k'_{0})^{2}\Big)
\Big((2 k_{1} -k'_{1})^{2} - A^{2}(2 k_{0}-k'_{0})^{2}\Big)}.
\ee
So far, this expression is exact, but now, we are going
to carry out the derivative expansion explained earlier,
which  coincides with the expansion in powers 
of  the momentun $k'$. The zeroth order term
in $k'$ contributes to the potential in $\chi$, and we
have explained above that if one starts with the correct
value of the expectation value of $\rho$, this term is
already taken care of. The first order term in $k'$
vanishes, and the second order term has a logarithmic
divergence. Setting
$$
I= I_{0}+ I_{1}+I_{2}+\ldots,
$$
$I_{1}$ is zero and $I_{2}$ has the form
\be
I_{2}= I_{2,0}\,(k'_{0})^{2}+ I_{2,1}\,(k'_{1})^{2},
\ee
and, after Euclidean rotation in $k_{0}$,
\bq
I_{2,0}&=&\frac{i}{2}\int d^{2}k \frac{k_{0}^{2}
(3 A^{2} k_{0}^{2} k_{1}^{2}+k_{1}^{4})}{(k_{1}^
{2}+A^{2} k_{0}^{2})^{4}},\nonumber\\
I_{2,1}&=&\frac{i}{2}\int d^{2}k \frac{k_{0}^{4}
(k_{1}^{2}- A^{2} k_{0}^{2})}{(k_{1}^{2}+ A^{2}
k_{0}^{2})^{4}}.
\eq

The integrals are elementary. After the change
of variables by
$$
k_{1}=r\sin(\theta),\;\;\; A \,k_{0}= r
\cos(\theta),
$$
the $\theta$ integrals are easily done,
leaving a logarithmically divergent $r$
integral:
\be
I_{2,0}=\frac{i\pi}{2 A^{3}}\int_{\epsilon}
^{\lambda} \frac{dr}{r},\;\;\;
I_{2,1}= -\frac{i\pi}{2 A^{5}}\int_{\epsilon}
^{\lambda}\frac{dr}{r}.
\ee
We have introduced both an infrared cutoff
$\epsilon$ and an ultraviolet cutoff $\lambda$
to regulate the divergent $r$ integral. Putting
this back into the equation for $S'_{2}$ (eq.()),
and rewriting it in the position space, we have
\be
S'_{2}\cong \frac{D\,R}{32\pi A}\int d\sigma
\int d\tau\left(A^{2} (\partial_{\tau}\chi)^{2}
-(\partial_{\sigma}\chi)^{2}\right),
\ee
where we have defined
$$
R=\int_{\epsilon}^{\lambda}\frac{dr}{r}.
$$

We close this section with a couple of comments
on this result:\\
a) The contribution of the $\chi$ mode to the action
is exactly the same as the contribution of ${\bf q}$
to $S_{m}$ (eq.(48)), apart from an overall cutoff
dependent multiplicative
constant. This constant can be eliminated by rescaling
$\chi$ (wave function renormalization). The important
point is that it is positive; otherwise, this term
would have the wrong (ghostlike) signature.\\
b) There is therefore an additional string mode
represented by $\chi$, with the same slope
$$
\alpha'= A/2
$$
as the other modes. This changes the dimension of
the target space from $D$ to $D+1$, and therefore
it can regarded as a non-leading contribution in the
large $D$ limit.\\
c) If we continued the derivative expansion of
$S'_{2}$ beyond second order, we would get terms with
higher derivatives of $\chi$ with respect to $\tau$
and $\sigma$. By simple power counting, these terms
would be finite and therefore they would not be of
interest to us.\\
d) Let us now consider terms higher then second
order in $\chi$ in the expansion of $S'$. The
logarithmically divergent contribution still comes
from second order in the derivative expansion.
Consider the graph of Fig.9, where two
external lines carry finite momenta and the rest
of the external carry zero momentum. A
simple generalization of the calculation given
above in the case of the second order term in
$\chi$ shows that this term must be of the form
\be
S'\cong \int d\sigma \int d\tau h(\chi)\left(
(\partial_{\tau}\chi)^{2} - A^{2} (\partial_
{\sigma}\chi)^{2}\right),
\ee
and $h(\chi)$ is a function whose power series
expansion in $\chi$ starts with the positive constant
that appears in eq.(80). Redefining the field $\chi$
by
$$
\chi\rightarrow g(\chi),
$$
and choosing the function $g$ so that it satisfies
$$
h(f(\chi)) (g'(\chi))^{2}=1,
$$
one can get replace $h$ by one. This shows that a simple
field redefinition gets us back to the second order result
given by eq.(80).

To complete our discussion, we should also consider
the contribution coming from $S_{g}$ in eq.(50). This
contribution has a quite different structure compared
to the one coming from the matter sector. It has a
linear cutoff dependence, and it depends only on
$\sigma$ derivatives of $\chi$ and not $\tau$
derivatives. We remind the reader of the
motivation behind
 integrating over the matter fields in $S_{m}$:
The integration produced a  kinetic energy term for the
originally non-propagating field $\chi$. Since
integrating over the ghost fields produces nothing
of comparable interest, it is probably best to
leave $S_{g}$ as it is.

\vskip 9pt

\noindent{\bf 9. Conclusions}

\vskip 9pt

This article is a direct follow up of reference [2].
The goal is to give a systematic treatment which
resolves various problems encountered in [2], while  still
staying within the framework of the simple meanfield
approximation developed in that reference. The 
crucial improvement over the treatment given in [2]
is the fixing of an accidental gauge invariance.
In addition, there are various other technical
improvements: Supersymmetry is introduced on the
world sheet to keep better track of matter-ghost
cancellation, and a better treatment of the 
singular determinants is presented. Also, following 
reference [4], we show how to impose partial Lorentz
invariance.

The meanfield method used here and in [2] is sufficiently
simple to allow the carrying out of a fairly complete
treatment. In the leading order of this approximation,
a condensation of Feynman graphs takes place, which
means that large order Feynman graphs dominate the
perturbation series. As a consequence, with a suitable
tuning of the coupling constant, a string with finite
slope is formed. We also show that a new dynamical degree
of freedom emerges, extending the transverse dimension
of the string from $D$ to $D+1$. Although the string
appears to be stable in the leading order, we argue
that the fundamental instability of the underlying
field theory probably shows up in the non-leading
orders of the approximation.

The question whether string formation takes place
in a given field theory is fundamentally important
but difficult to answer in practice.
The  meanfield approximation used in
this article, free from its earlier shortcomings, 
 is simple yet powerful enough to answer this
question in the case of the $\phi^{3}$ theory.
The possibility of applying this method to more
realistic theories appears more appears within reach.
Another valuable line of future research would be
to try to improve over the meanfield approximation. 

\vskip 9pt

\noindent{\bf Acknowledgements}

\vskip 9pt

This work was supported in part by the Director,
Office of Science, Office of High Energy and 
Nuclear Physics, of the US Department of Energy
under Contract DE-AC03-76SF00098, and in part
by the National Science Foundation under Grant
PHY-0098840.

\vskip 9pt

\noindent{\bf Appendix}

\vskip 9pt

In this appendix, we will discuss an alternative, and 
more conventional way of doing the mean field approximation,
and we will show that it is completely equivalent to the
approach used in this article. Instead of $\rho$ (eq.(39)),
we introduce a different set of composite fields:
\bq
\Delta S&=&\int d\sigma \int d\tau\Big(\kappa_{1}({\bf y}
\cdot \dot{{\bf q}} -\lambda_{1})+\kappa_{2}(\bar{{\bf b}}
\cdot {\bf b} -\lambda_{2})+\kappa_{3}(\bar{{\bf c}}\cdot
{\bf c}- \lambda_{3})\nonumber\\
&+&\kappa_{4}({\bf z}\cdot{\bf r} -\lambda_{4}) +\kappa_{5}
({\bf y}\cdot {\bf z}- \lambda_{5})+ \kappa_{6}(\bar{{\bf b}}
\cdot \bar{{\bf c}} -\lambda_{6})\Big).
\eq
In the large $D$ limit, the fields $\kappa_{i}$ and
$\lambda_{i}$ become classical and they can be replaced by
their ground state expectation values:
$$
\kappa_{i}\rightarrow \langle \kappa_{i}\rangle,\;\;\;
\lambda_{i}\rightarrow \langle \lambda_{i}\rangle.
$$
The justification for this is well known [16]: The
composite fields $\lambda_{i}$ are each sum of $D$ terms,
and consequently, they grow like $D$ as $D$ becomes large.
On the other hand, compared to this,
the quantum fluctuations are suppressed by a factor of
$1/\sqrt{D}$.
 As a result, in leading order large $D$ these
fields become classical. In contrast, there is no comparable direct
argument for why $\rho$  should become classical
in this limit. 
 Therefore, from the perspective
of large $N$ (large $D$) physics, the approach sketched
in this appendix is better motivated than the approach
developed in the main body of the article. The disadvantage
 is that, at least
initially, many more auxilliary fields have to be introduced.
We will show presently that the expectation values of
 these extra fields can easily be
expressed in terms $\kappa_{0}$ and $\rho_{0}$ introduced
earlier. To do this, we first write $S_{f}$, the fermionic
part of the action, in terms of the fields introduced in
eq.(82): 
\bq
S_{f}&=& \int d\sigma \int d\tau\Big(i\bar{\psi} \dot{\psi}
+\frac{g}{p^{+}}(\bar{\psi}_{1}\bar{\psi}_{2}+\psi_{2}
\psi_{1})\nonumber\\
&+&\frac{1}{2}(\lambda_{1}+\lambda_{2}+\lambda_{3}
+p^{+} \lambda_{4}-\alpha \lambda_{5}-\alpha \lambda_{6})
\sum_{i}\bar{\psi}_{i}\psi_{i}
+\frac{\alpha}{a}(\lambda_{5}+\lambda_{6})\Big),\nonumber\\
& &
\eq
where, to keep things simple, we have set all the $\beta$'s equal
to one. Comparing this with the $S_{f}$ given by eq.(41) leads
to the identification
\be
\kappa_{0}=\lambda_{1}+\lambda_{2}+\lambda_{3}+p^{+}
\lambda_{4}-\alpha \lambda_{5} -\alpha \lambda_{6},
\ee
so that the two expressions for $S_{f}$, apart from
an additive term, agree.
We now consider all the $\lambda$ dependent terms in
$$
\Delta S+S_{f}
$$
and write down the equations of motion by varying
the $\lambda$'s, subject, however, to the constraint
that $\kappa_{0}$ (eq.(84)) is held fixed. These equations
give
\be
\kappa_{1}=\kappa_{2}=\kappa_{3}=\frac{\kappa_{4}}{p^{+}},
\;\;\;\kappa_{5}=\kappa_{6}=\alpha(\frac{1}{a}-\kappa_{1}).
\ee
Consequently, there is only one independent $\kappa$, say
$\kappa_{1}$. With the further identification
\be
\kappa_{1}=\rho_{0},\;\;\;\frac{1}{a}-\kappa_{1}=\bar{\rho}_{0},
\ee
we recover the action of eq.(36) with $\beta$'s set equal to one,
 establishing the equivalence
of the two approaches.

\vskip 9pt

{\bf References}
\begin{enumerate}
\item K.Bardakci and C.B.Thorn, Nucl.Phys. {\bf B 626} (2002) 287,
hep-th/0110301.
\item K.Bardakci and C.B.Thorn, Nucl.Phys. {\bf B 652} (2003) 196,
hep-th/0206205.
\item K.Bardakci and C.B.Thorn, Nucl.Phys. {\bf B 661} (2003) 235,
hep-th/0212254.
\item K.Bardakci, Nucl.Phys. {\bf B 667} (2004) 354, hep-th/0308197.
\item C.B.Thorn, Nucl.Phys. {\bf B 637} (2002) 272, hep-th/0203167.
\item S.Gudmundsson, C.B.Thorn and T.A.Tran, Nucl.Phys. {\bf B 649}
(2003) 3, hep-th/0209102.
\item C.B.Thorn and T.A.Tran, Nucl.Phys. {\bf B 677} (2004) 289,
hep-th/0307203.
\item G.'tHooft, Nucl.Phys. {\bf B 72} (1974) 461.
\item A.Clark, A.Karsch, P.Kovtun and D.Yamada, Phys.Rev.
{\bf D 68} (2003) 066011.
\item R.M.Koch, A.Jevicki and J.P.Rodrigues, Phys.Rev. {\bf D 68}
(2003) 065012, hep-th/0305042.
\item O.Aharony, J.Marsano, S.Minwalla, K.Papadodimas and
M.V Van Raamsdonk, hep-th/0310285.
\item R.Gopakumar, hep-th/0402063.
\item J.M.Maldacena, Adv.Theor.Math.Phys.{\bf 2} (1998) 281,
hep-th/9711200.
\item O.Aharony, S.S.Gubser, J.Maldacena, H.Ooguri and
Y.Oz, Phys.Rep. {\bf 323} (2000) 183, hep-th/9905111.
\item P.Goddard, J.Goldstone, C.Rebbi and C.B.Thorn,
Nucl.Phys. {\bf B 56} (1973) 109.
\item For a recent review of the large N method, see
M.Moshe and J.Zinn-Justin, Phys.Rept. {\bf 385}
(2003) 69, hep-th/0306133.
\item A.D'Adda, M.Luscher and P.Di Vecchia, Nucl.Phys.
{\bf B 146} (1978) 63. 
\item D.Gross and A.Neveu, Phys.Rev. {\bf D 10} (1974)
3235.
\end{enumerate}

\end{document}